\documentclass{llncs}

\usepackage{blindtext, graphicx, subfigure, url}

\let\llncssubparagraph\subparagraph
\let\subparagraph\paragraph
\usepackage[compact]{titlesec}
\let\subparagraph\llncssubparagraph

\usepackage{booktabs} 
\usepackage{array}
\newcolumntype{L}{>{$}l<{$}}
\newcolumntype{C}{>{$}c<{$}}
\newcolumntype{R}{>{$}r<{$}}

\usepackage{multirow} 

\usepackage{cite}   
\usepackage[square, numbers, comma, sort&compress]{natbib}  

\makeatletter
\renewcommand\bibsection%
{
  \section*{\refname
    \@mkboth{\MakeUppercase{\refname}}{\MakeUppercase{\refname}}}
}
\makeatother

\newcommand{\eg}{e.g.,\xspace}
\newcommand{\ie}{i.e.,\xspace}

\newcommand{\etc}{etc.\@\xspace}

\usepackage[title=normal,margins=normal,indent=normal,bibliography=normal,bibnotes=normal]{savetrees}

\begin{document}

\title{Characterizing BigBench queries, Hive, and Spark in multi-cloud environments}

\author{Nicolas Poggi, Alejandro Montero, David Carrera}

\institute{Barcelona Supercomputing Center (BSC)\\
Universitat Polit\`{e}cnica de Catalunya (UPC-BarcelonaTech)\\
 Barcelona, Spain}
\maketitle

\begin{abstract}

BigBench is the new standard (TPCx-BB) for benchmarking and testing Big Data systems. The TPCx-BB specification describes several business use cases ---queries--- which require a broad combination of data extraction techniques including SQL, Map/Reduce (M/R), user code (UDF), and Machine Learning to fulfill them.  However, currently, there is no widespread knowledge of the different resource requirements and expected performance of each query, as is the case to more established benchmarks. 
Moreover, over the last year, the Spark framework and APIs have been evolving very rapidly, with major improvements in performance and the stable release of v2.  It is our intent to compare the current state of Spark to Hive's base implementation which can use the legacy M/R engine and Mahout or the current Tez and MLlib frameworks.

At the same time, cloud providers currently offer convenient on-demand managed big data clusters (PaaS) with a pay-as-you-go model. In PaaS, analytical engines such as Hive and Spark come ready to use, with a general-purpose configuration and upgrade management.   The study characterizes both the BigBench queries and the out-of-the-box performance of Spark and Hive versions in the cloud.  At the same time, comparing popular PaaS offerings in terms of reliability, data scalability (1GB to 10TB),  versions, and settings from Azure HDinsight, Amazon Web Services EMR, and Google Cloud Dataproc.  The query characterization highlights the similarities and differences in Hive an Spark frameworks, and which queries are the most resource consuming according to CPU, memory, and  I/O.
Scalability results show how there is a need for configuration tuning in most cloud providers as data scale grows, especially with Sparks memory usage.  These results can help practitioners to quickly test systems by picking a subset of the queries which stresses each of the categories. At the same time, results show how Hive and Spark compare and what performance can be expected of each in PaaS.

\end{abstract}

\section{Introduction}

A benchmark captures the solution to a problem and guides decision making.
For Big Data Analytics Systems (BDAS) BigBench has been recently standardized by the Transaction Processing Performance Council (TPC) as TPCx-BB. It has been originated from the need to expand previous decision support style benchmarks i.e., TPC H~\cite{TPC-H} and DS~\cite{TPC-DS} into semi and non-structured data sources, and it is the result of many years of collaboration between the database industry and academia~\cite{BigBench,BigBench2TPCxBB}.
BigBench includes 30 business uses cases ---queries--- covering merchandising, 
pricing optimization, product return, and customer questions.  It's implementation requires a broader set of data technologies than SQL \ie Map/Reduce (M/R), user code (UDF), Natural Language Processing (NLP), and Machine Learning; which expand and differentiates from previous SQL-only benchmarks, as required by today's enterprise.  

BigBench's original implementation was based on Apache Hadoop and Hive with Map/Reduce (M/R) as execution engine and Mahout as Machine Learning (ML) library~\cite{BigBench2TPCxBB}.  However, due to the rapid development of the open source Big Data ecosystem and BigBench online repository~\cite{BigBench_github}, it is now possible ---and convenient--- to use \eg Tez as execution engine and MLlib to lower the query latency, or even to replace Hive altogether with Spark.  However, there is not only a large set of technologies that a user can choose from, but there are multiple stable major versions of the frameworks.  Moreover, Spark ---and MLlib--- have been evolving rapidly over the last year, with major enhancements in performance and API changes with v2.  It is impart our intent to quantify and understand the performance improvements of such changes from the base implementation.  

At the same time in recent years, new managed enterprise big data services have emerged in most cloud providers~\cite{ForresterPaaS}, facilitating software-defined on-demand big data deployments. 
These services create compelling technical reasons for migration to the cloud, such as elasticity of both compute and storage, while maintaining a simplified infrastructure management i.e. via \textit{virtualization}. Furthermore, with such services often using a \emph{Pay-as-you-Go} or even \emph{Pay-as-you-Process} pricing model, they are economically attractive to customers~\cite{ForresterPaaS}. 
Furthermore, cloud providers make the complex configuration and tuning process~\cite{BD14} transparent to their clients, while providing features such as data security and governance. On top of this, by having multiple customers, service providers can potentially improve their software-stack from user feedback, as upgrading services more often than smaller companies~\cite{BD16}. As a result, the client can benefit from the immediate availability of a tested and generically-optimized platform with upgrade management.

The current cloud and open source Big Data ecosystem, leaves the enterprise facing multiple  decisions that have can an impact both in the budget as in the agility of their business.  These include selecting both an infrastructure and services provider, as well as the Big Data frameworks along their configuration tuning~\cite{BD14}. With this respect, BigBench becomes the clear choice to contrast cloud providers and choose the appropriate data framework to make an appropriate choice.  However, as being a new benchmark, still little is understood of the underlying implementation and expected performance of the queries.  To day, only a handful of official submissions are available~\cite{BigBench_submissions}, as well as a few publications with detailed per query characterization~\cite{Ivanov2016,BigBench2TPCxBB}.

The goal of this study is two-fold. First, it provides a first approach to BigBech query characterization to understand both Hive and Spark implementations. Second, it compares the out-of-the-box performance and data scalability from 1GB to 10TB of popular cloud PaaS solutions. Surveyed services include Azure HDinsight, Amazon Web Services EMR, and Google Cloud Dataproc, as well as an on-premises cluster as baseline.  The work is the natural extension of the previous cloud SQL-on-Hadoop comparison, where the same cloud providers where compared using the TPC-H SQL benchmark~\cite{BD16} and only using Hive.  The reason for not using Spark before, was that the versions, performance and configuration were not stable or comparable enough among providers. The study targets medium size clusters of 128-core each. In particular, we benchmark similar clusters in each provider consisting of 16 data nodes, with a total of 128 worker CPU cores and about 60GB of RAM, using networked/block storage only.  The master nodes where chosen with 16-cores each and more than 60GB of RAM to sustain the concurrency tests.  

Objectives and contributions:

\begin{enumerate}
	\item A characterization of the different BigBench queries to better understand the use cases resource requirements and implementation differences of Hive and Spark.
	\item Survey the popular entry level PaaS Hadoop solutions from main cloud providers and contrast offerings and data scalability using networked/block storage.
	\item Compare the performance of the different versions of the Big Data Apache/Hadoop ecosystem, as well as the machine learning libraries.	
\end{enumerate} 

\subsubsection{Organization}
The rest of the study is organized as follows.  
Section 2 presents the cloud providers and cluster hardware and software specs.  
Section 3 presents the background of the different technologies as well as the most relevant state-of-the art in the field.    
Section 4 presents the methodology used for the testing as well as the query characterization by resource usage for both Hive and Spark.
While Section 5 presents the main query performance results at different data scales.
Section 6 adds to discussion other tests performed with different versions of the frameworks. While Section 6 discusses the results and Section 7 provides a summary and the conclusions. 

%
%


\section{Providers and Systems-Under-Test (SUTs)}
\label{sec:SUTs}

The Hadoop PaaS services from 3 major cloud providers are compared:

\begin{itemize}
	\item HDInsight (HDI) from Microsoft Azure. 
	\item Elastic Map Reduce (EMR) from Amazon Web services (AWS). 
	\item Dataproc (GCD) from Google Cloud Platform. 
\end{itemize}

The elasticity properties of the 3 providers have been previously studied~\cite{BD16} using a derived TPC-H benchmark implemented in Hive~\cite{D2F}, along with Rackspace's Cloud Big Data (CBD) not included in this report.
There were different reasons justifying the selection of each provider. HDI has been studied previously~\cite{BD15,BD14}, and as such, their APIs are already well-integrated into the ALOJA platform (see Section~\ref{sec:background}). EMR was the first major Hadoop PaaS solution, and currently has one of the largest usage shares~\cite{ForresterPaaS}. Both AWS and EMR are commonly used as bases of comparison in the literature~\cite{Zhang2014cloud, zhang2014optimizing, SruthiEMR}. GCD from Google Cloud has been included due to it being identified as a leading provider~\cite{ForresterPaaS}, as well as for being a new service (GA in 2016) which could potentially have a differentiated architecture.

\subsubsection{HDInsight} for HDI, we have used the D4v2 VM instances, which features 8-cores and 28GBs of RAM.  The HDFS on all HDI instances is backed by the Azure Blob store (through the WASB driver). This means that it is an object-store over the network. As a consequence, the storage on HDI is decoupled from compute and can grow elastically, as well as be used from outside the HDFS cluster on other shared services and users. Local disks, backed by SSD drives on the D-series, are \emph{ephemeral} and used for temporary or intermediate data only.  The two included master nodes are of the D14v2 instance type, featuring 16-cores each and 112GB of RAM. Deployment times in Azure took close to 20 minutes on most builds.  The on-demand price for the cluster was  \$20.68 per hour, billed by the minute.

\subsubsection{Elastic Map Reduce} for EMR, the default m4.2xlarge instance was tested.  Is is an EBS-only instance. It comes with 8-cores and 32GB of RAM. EBS stands for Elastic Block Store, Amazon's over-the-network storage. EBS has 4 different throughput (IOPS) plans, according to the technology backing the storage. The plans being high-performance or regular, for both SSDs and rotational drives; we chose the default regular SSDs (GPS2).  The master node was the m4.4xlarge with 16-cores and 64GB of RAM. Deployment times were faster than HDI at around 10 minutes.  ERM is billed by the hour or fraction of hour, being the only provider maintaining this high billing. The cluster on-demand  price was of \$10.96 per hour.

\subsubsection{Cloud Dataproc} for GCD, we have evaluated the n1-standard-8  instance with 8-cores and 30GB RAM, with the Google Cloud Storage (GCS)---the network based storage.  In GCD, up to 4 SSDs can be added per node at creation time and the volumes are not \emph{ephemeral} as in HDI, but used for HDFS. Deployment times were surprisingly fast for GCD, with cluster build times at around 1 minute.  The master node was of the n1-standard-16 with 16-cores and 60GB of RAM, GCD has the option to include one or two master nodes, we choose one. GCD is billed by the minute.   The cluster had an on-demand cost of \$10.38 per hour.

\subsection{Software stack and versions}
While EMR, dating back to 2009, was the first major PaaS Hadoop solution, the other main providers have caught up in packaging Hadoop with other popular ecosystem services. Currently, all of the four providers tested ships with both \textit{Hive} and \textit{Spark} v2 for SQL-like analysis on top of Hadoop, as well as other services and tools \ie Presto, HBase, Storm, Pig, \etc. There are differences in security and governance features, but these are not compared in this work.
In relation to the software versions, Azure base their PaaS solutions on the Hortonworks Data Platform (HDP)~\cite{HDP}, a popular Hadoop distribution that users might already be familiar with.  HDI uses Ubuntu Linux 16.04 as Operating System (OS). During the main experimental period---June 2017---HDI added support for HDP version 2.6, which features Spark 2.1, but still runs on Hive v1.
AWS uses a custom-built stack for EMR, as well as a custom Linux version called the Amazon Linux AMI, in this case v 2017.03.  The EMR tests were run with the latest version available at the time of testing, EMR 5.5 featuring both Hive and Spark at v2.   
Like AWS, Google's Cloud Dataproc also uses a custom Hadoop distribution built using Apache BigTop (as EMR also does), and Debian Linux 8.4 OS. Tests were run on the \textit{preview} version, as the current v1.1 only featured Spark 2.0, and we wanted to test the 3 of them with Spark 2.1.  A comparison among Spark versions 1.6.3, 2.0, and 2.1 is provided in Section~\ref{sec:additional}.
Versions tested for the SUTs are the default ones offered at time of cluster deployment during June 2017. More information on the software versions can be found on the release notes of each provider.
Relating to data center zones, we have tested HDI at \emph{South Central US}, EMR at \emph{us-east-1}, and GCD at \emph{west-europe-1}.

\subsubsection{Software configuration differences}
\label{sec:SW-HW_conf}

\begin{table*}[]
	\centering
	\caption{Most relevant Hadoop-stack configurations for providers }
	\label{tab:config}
	\resizebox{\textwidth}{!}{%
	\begin{tabular}{@{}lllllll@{}}
		\toprule
		\textbf{Category}                & \textbf{Config}          & \textbf{EMR}       & \textbf{HDI}       & \textbf{GCD}      \\ \midrule
		\multirow{2}{*}{\textbf{System}} & OS                       & Linux AMI 2017.03      & Ubuntu 16.04       & Debian 8.4         \\
		& Java version             & OpenJDK 1.8.0\_121  & OpenJDK 1.8.0\_131 & OpenJDK 1.8.0\_121  \\ \midrule
		\multirow{4}{*}{\textbf{HDFS}}   & File system              & EBS        & WASB               & GCS                       \\
		& Replication              & 3                  & 3                  & 2                 \\
		& Block size               & 128MB              & 128MB              & 128MB            \\
		& File buffer size         & 4KB                & 128KB              & 64KB             \\ \midrule
		\multirow{3}{*}{\textbf{M/R}}    & Output compression       & SNAPPY             & FALSE              & FALSE                  \\
		& IO Factor / MB           & 48 / 200            & 100 / 614          & 10 /100           \\
		& Memory MB                & 1536               & 1536               & 3072             \\ \midrule
		\multirow{2}{*}{\textbf{Hive}}   & Hive version                   & 2.1                 & 1.2                & 2.1                  \\
		& Engine                   & Tez                 & Tez                & M/R                 \\ \midrule

		\multirow{6}{*}{\textbf{Spark}}   & Spark version                   & 2.1.0.2.6.0.2-76                 & 2.1                & 2.1.0                  \\
		& Driver memory                   & 5G                 & 5G                & 5G                 \\
		& Executor memory               & 4G           & 5G           & 10G           \\
		& Executor cores          & 4              & 3            & 4               \\
		& Executor instances            & Dynamic              & 20            & Dynamic               \\
		& dynamicAllocation  enabled        & TRUE              & FALSE               & TRUE                   \\
		& Executor memoryOverhead & Default (384MB)                &    Default (384MB)            & 1,117 MB                \\ 
		
		\bottomrule
	\end{tabular}
}
\end{table*}

While this work focuses on the out-of-the-box performance of PaaS services, the difference in execution times from Section~\ref{sec:times} for SUTs with similar HW led us to compare configuration choices in detail as summarized by Table~\ref{tab:config}.  Note that while the Java versions are 1.8, all providers used the OpenJDK versions, as opposed to Oracle's JDK as traditionally recommended. At the HDFS level, all providers used their object networked-based storage.  As object stores are typically replicated besides by Hadoop, it was interesting to see that only GCD and CBD lowered the HDFS replication to 2 copies.  Most block sizes were the default at 128MB, while only EMR used the default \emph{file buffer sizes}.  EMR and CBD both compressed the map outputs by default, and each tuned the I/O factor and the I/O MBs.

While there were some differences at the Java/Hadoop level, tuning Hive had the most significant impact on performance. It was remarkable that GCD did not have the Tez execution engine enabled by default. Tez reduces the total BigBench running time by 2-3x as shown in Section~\ref{sec:legacy}.  There were also differences in other Hive parameters among providers, such as using the \textit{cost-based optimizer}, \textit{vectorized execution} and bucket settings.  The provider that enabled most performance improvements in Hive \ie HDI, got the best results for similar HW.  
In the case of Spark, the main results all use versions 2.1.  The main difference among providers is that GCD has twice the memory per executor, at 10GB.  Another interesting difference is that HDI is still setting the number of executor instances statically, while both EMR and GCD use dynamic allocation.  Dynamic allocation is a recent feature that we are exploring on our on-premises cluster currently to quantify the trade offs.

\section{Background and Related work}
\label{sec:background}

The motivation behind this work is to expand the cloud provider survey~\cite{BD16} from SQL-only TPC-H using Hive into BigBench.  As well as expanding the work to include Spark and update the results to the current versions and larger cluster sizes ---from 8 to 16 datanodes. The work is integrated into the ALOJA benchmarking and analysis platform~\cite{BD14,BD15}, which can be used to reproduce the tests and provide access to the raw results from these tests. The ALOJA project is an open initiative from the Barcelona Supercomputing Center (BSC) to explore and automate the characterization of cost-effectiveness for big data deployments.
Furthermore, in a previous study with Ivanov et. al., in collaboration with the SPEC research group in big data benchmarking~\cite{SPECBD} led to the generation of a ``Big Data Benchmark Compendium''~\cite{Compendium}, which surveys and classifies the most popular big data benchmarks. This work represents also an expansion of the survey by adding BigBench results for Hive and Spark. 

%
%

\subsubsection{TPCx-BB (BigBench)~\cite{BigBench}}
\label{sec:BB}
BigBench is an end-to-end application level benchmark specification standardized by TPC as TPCx-BB. It is the result of many years of collaboration between industry and academia. Covering most Big Data Analytical properties (3Vs) in 30 business use cases ---queries--- for a retailer company in the areas of merchandising, 
pricing optimization, product return, and customers. 
It is also able to scale the data from 1GB to petabytes of data.
The BigBench v1.2 reference Implementation resulted in:
\begin{itemize}
	\item 14 declarative queries (SQL): 6, 7, 9, 11, 12, 13, 14, 15, 16, 17, 21, 22, 23, 24.
	\item 3 with user code (UDF): 1, 29, 30 (also uses M/R).
	\item 4 Natural Language Processing (NLP) 10, 18, 19, 27.
	\item 4 with data preprocessing with M/R jobs: 2, 3, 4, 8.
	\item 5 with Machine Learning jobs: 5, 20, 25, 26, 28. 
\end{itemize}

%
%
%
%
%
%



\subsubsection{Apache Hive~\cite{Hive}} has become the \emph{de facto} data-warehousing engine on top of Hadoop, providing data summarization, analysis, and most importantly, support for SQL-like queries. The Hive engine can be controlled using HiveQL, an SQL-like language designed to abstract the Map/Reduce (M/R) jobs involved in such queries for analysts. As an example, Hive queries have been gradually replacing the use of M/R in large companies such as Facebook and Yahoo!~\cite{Yahoo}. 
In Hive the default engine to manage task executions is Hadoop's M/R.  However, in the latest versions Hive added support for different execution engines. Namely Tez~\cite{Yahoo} (from the Stinger project) is a popular drop-in replacement, which improves on the M/R model to provide lower latency and performance.  The Hive configuration employed by the different SUTs is described further in Section~\ref{sec:SW-HW_conf}.  For ML, Hive relied originally on the Mahout library, however, it can also use the MLlib provided by Spark more actively developed as use in the main results of this work.

\subsubsection{Apache Spark~\cite{Spark}} similar to Hive, Spark is another popular framework gaining momentum~\cite{Ivanov2016}. Spark is a processing engine which provided increased performance over the original Map/Reduce by leveraging in-memory computation. 
Spark was originally created in 2009 by the AMPLab at UC Berkeley and was developed to run independently of Hadoop. The Spark project consists of several integrated components including: the Spark core as general execution engine and APIs,  Spark SQL for analyzing structured data, Spark Streaming for
analyzing streaming data, MLlib for Machine Learning, and GraphX for graph analytics.  During the past year, the Spark API had suffered significant changes and improvements, many in the area of performance.  Now stabilizing at v2, and with most cloud providers supporting the current versions, Spark becomes a natural choice as an integrated framework over Hive and Mahout.

\subsection{Related work}
Most recent evaluations of Hadoop systems in the cloud are SQL-only~\cite{Floratou2014,BD16}. This is the first attempt to measure performance of cloud-based SUTs and make comparisons between the main providers using BigBench, which expands use cases from the SQL-only boundary and includes both Hive and Spark v2 results.
There are already several tests of Amazon's EMR services in the literature: Sruthi~\cite{SruthiEMR} presents the performance and costs models of EMR (PaaS) vs. AWS (IaaS) using the Intel HiBench~\cite{HiBench} benchmark suite, but only includes a minimal Hive-test benchmark based on Pavlo's CALDA\cite{Pavlo2009} benchmark, concluding that PaaS solutions benefit from being provider-optimized. Zhang et. al. ~\cite{Zhang2014cloud, zhang2014optimizing} focuses on scheduling of jobs in EMR and on Hadoop configuration using micro-benchmarks similar to our previous work~\cite{BD14}. 
In~\cite{Floratou2015} Floratou et. al. describe the current problems associated with benchmarking SQL-on-Hadoop systems, and advocate the standardization of the process.  We believe that BigBench~\cite{BigBench} is the current reference benchmark for such systems.   Relating to BigBench query results, to day there are only a handful of official submissions are available~\cite{BigBench_submissions}, as well as a few publications with detailed per query characterization~\cite{Ivanov2016,BigBench2TPCxBB}.
More established benchmarks \ie TPC-H have been thoroughly analyzed, in work including the query their \emph{choke points} as in ``TPC-H Analyzed"~\cite{Boncz2014}. It is the intent to provide here a first look into BigBenche's \emph{choke points}. This work expands on the available BigBench results by including Spark v2, along with detailed per query characterization, and cloud data scalability results.

\subsection{Legacy execution engine and Machine Learning frameworks}
\label{sec:legacy}
\begin{figure*}[!ht]
	\centering
	\subfigure[Map/Reduce vs Tez]{
		\includegraphics[width=0.5\textwidth]{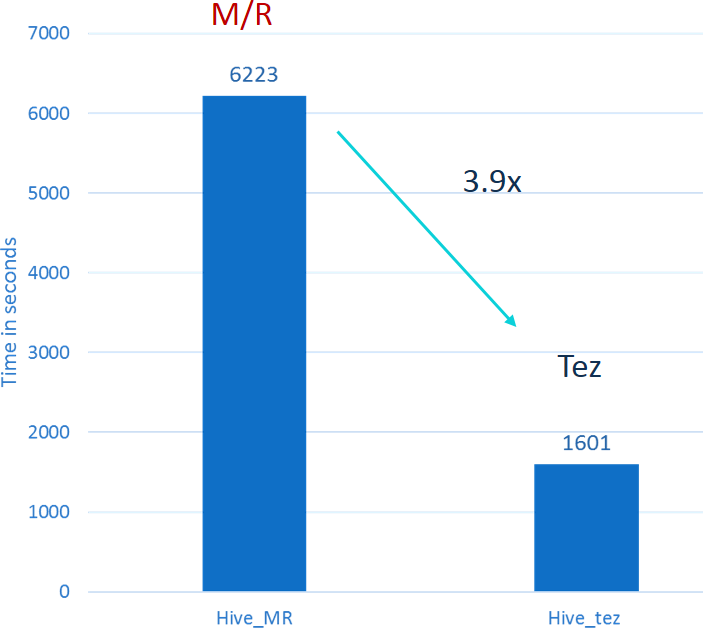}
		\label{fig:MR}		
	}
	~
	\subfigure[Mahout vs. MLlib v1]{
		\includegraphics[width=0.33\textwidth]{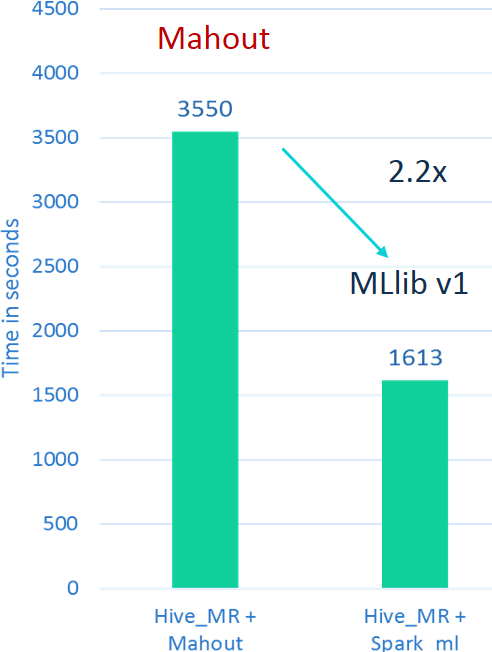}
		\label{fig:ML}		
	}
	\caption{Comparison of legacy vs. current Hive execution engine and ML library at 100GB}
	\label{fig:legacy}
\end{figure*}

In a previous work~\cite{HSEU2017EU}, we have compared the performance of Hive with M/R vs. Tez as execution engine, as well as Mahout vs. MLlib v1.
Figure~\ref{fig:MR} shows preliminary results comparing the Map/Reduce (M/R) execution engine to Tez in Hive at 100GB. While Figure~\ref{fig:ML} shows Hive using the Mahout Machine Learning library vs. the newer MLlib from the Spark package.
It can be seen that Tez can be up to 3.9x faster than the classical M/R.  For this reason, all of the following tests are done with Tez when possible.  Similarly, for Machine Learning, MLlib v1 from the Spark package can be at least twice as fast than the now legacy Mahout framework.  
For more details refer to ~\cite{HSEU2017EU}.

\section{Methodology and Query characterization}
\label{sec:methodology}
The configuration of the tested PaaS services is left as the provider pre-configures them, so that the out-of-the-box performance can be measured without introducing bias towards a provider.  This means that the HDFS configuration, default execution engine i.e. M/R or Tez, Hive, and OS are all left unmodified. Our intentions are not to produce the best results for the hardware as in our previous works~\cite{BD14,BD15}, but instead to survey different instance combinations, and test the clusters as the provider intended for general purpose usage. We also expect each provider to optimize to their respective infrastructures, especially to their storage services. Where information \ie underlying physical hardware, might not be readily available to implement in a custom IaaS fashion. By using defaults, the aim is to provide an indication of experience that an entry-level user might be expected to have on a given system without having to invest in additional fine-tuning.
Specific configurations can be found at Section~\ref{sec:SW-HW_conf} and at the providers release page for further reference.

As test methodology, the 30 BigBench queries are run sequentially and at least 3 times for each SUT and scale factor, capturing running times. We test each SUT with scale factor 1, 10, 100, 1000 (1TB), and 10000 (10TB).  
The metrics we present are the execution times of each individual query or in together in the concurrency tests. Opposed to the final BigBech queries per minute (BBQpm) metric, only permitted in audited results.
The tests were run in June 2017, and the reported pricing and specs correspond to this period.  All of the instances are using the on-demand (non-discounted) pricing, which usually is the highest, but simplifies calculations. 
As a side note, prices should only be regarded as indicative; in~\cite{BD16} we have noted that in a 4-month test period, prices were changed at least 2 times for some providers, while also new versions of the software stack were released.



%

\subsection{Query Characterization}

\begin{figure}[]
	\centering
	\centerline{ 
	\includegraphics[width=1.45\textwidth]{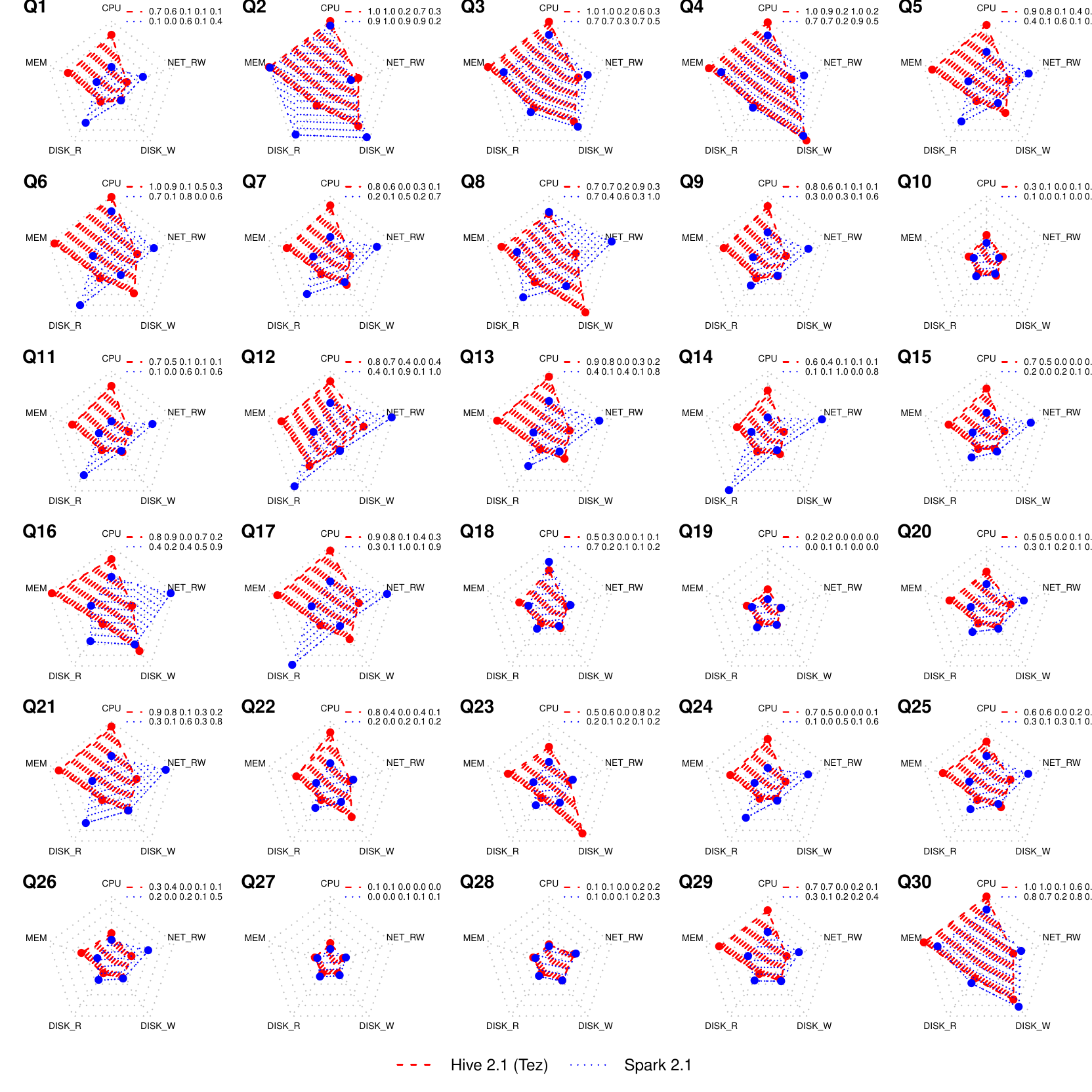}
   }
	\caption{Per query resource consumption radar chart of BigBench at 1TB for Hive and Spark. Legend prints CPU, RAM, Disk R/W, and network normalized resource usage respectively.}
	\label{fig:char_query}
\end{figure}

As BigBench is a recent benchmark with constant changes and additions to the implementation, there is still little knowledge about the difference in queries besides the official submitted results~\cite{BigBench_submissions} and the works of~\cite{Ivanov2016,BigBench2TPCxBB}. Per query knowledge can be useful to quickly benchmark a new SUT \eg by \emph{cherry-picking} queries that stresses different resources, or that are the most demanding ones.  In this way, a user can perform a \emph{smoke test}, or spot configuration problems more rapidly. Also, this knowledge is useful to validate results, optimize the current implementation, or when adding new systems as in more established benchmarks~\cite{Boncz2014}.  

Figure~\ref{fig:char_query} presents a radar chart of the 30 BigBench queries by normalized system resource utilization, comparing Hive and Spark EMR at 1TB.  The system resources requirements represent the query's average utilization so that measurements are not dominated by query times. Results are normalized between 1 (maximum) and 0 (minimum) as they have different units.  They include 5 separate metrics including CPU percentage utilization, MEM for main memory (RAM), Disk Read, Disk Write, and Network R/W.  The network resource averages the R/W as the traffic is only internal and both values aggregate to the same amount. To obtain these numbers, performance metrics where captured in ALOJA using the \textit{sysstat} package.

The chart can be read as follows: each query, in order is presented in its own subplot;  each radar chart has the 5 resource dimensions (read counter-clockwise); the larger the area, the more resources consumed; Hive in red long dashes and Spark in blue short dashes.  Additionally, the small legend on the upper left corner prints the values of each resource in order, and we be used to export the results (due to space limitations).  The EMR SUT was selected to represent the resource consumption due to GCD using M/R, and as we saw in Section~\ref{sec:legacy} the performance is not comparable to Tez. HDI was discarded as access to the storage is counted as network traffic, opposed to EMR and GCD. This resulted in the chart representing Hive 2.1 and Spark 2.1, both with MLlib v2.  

Figure~\ref{fig:char_query} shows visually the highest resource demanding queries, as well as the similarities and differences between Hive and Spark.  The list of query categories can be found in Section~\ref{sec:BB}.  The queries with the highest resource usage in both systems are the M/R style of queries.  Query 2 is the highest for Spark and second for Hive, while Q4 is the highest for Hive and second for Spark.  Besides Q2 and Q4, Q30 is the next with highest resource requirements for both systems, while Q30 is in the UDF category, it also makes use of M/R style processing.  So in general, the M/R queries need the highest resources and similar behavior in Hive and Spark.  In contrast, the queries with the lowest requirements are the Natural Language Processing (NLP) group.  They are implemented for both systems using the OpenNLP framework as UDFs.

The SQL-type queries is the second group with the highest requirements for some queries ---also the most numerous, so they cover a wide range of values. Query 6 has the highest requirement on Hive ---but not in Spark, while Q16 is the highest in Spark and second in Hive.  In general, the SQL-type queries are the most dissimilar between Hive and Spark. Hive+Tez in general have a very high CPU requirement, while Spark reads more in average from the disks.  This read from disk difference is particularly high for Q14 and Q17, the queries which requires a cross-join which is discussed further on Section~\ref{sec:errors} as limitation in Spark.

\subsection{Hive vs. Spark approaches}
Comparing Hive to Spark, their configuration and execution approach in the 3 providers is quite distinct.  While Hive and Tez are configured to use many but small containers, Spark is configured the opposite way.  Spark favors \emph{fat} containers, fewer containers, but with more CPUs and memory each.  While Hive is typically setup in the traditional M/R way of independent and redundant containers, which could be potentially more scalable and fault-tolerant, but also requires higher disk writes as shown in Figure~\ref{fig:char_query}. Tez also has a higher CPU utilization than Spark, which is not using all of the cluster resources when running single queries in the power tests. Spark in general has a higher network usage (shuffling data) and more intensive in disk reads.  Surprisingly it uses less memory than Hive, but this is explained as not all of the available containers per node are used in single query runs.

Even though the different configuration approaches, about half of the queries have similar behavior in both systems.  These queries being: 2, 4, 3, 10, 18, 19, 20, 22, 25, 26, 27, 28, 29, and 30; while queries 8, 9, 13, 15, 16, 21, 23 share some similarities.  As mentioned, the SQL queries are the most differentiated, including queries: 1, 5, 6, 7, 11, 12,  14, 17, 24.

In the ML category, while the query times are significantly different, Hive is also using the MLlib v2 library from Spark. So both systems have very similar resource requirements.  The ML query with the highest resource needs is Q5, which performs a logistic regression.  Query 5 is analyzed in more detail bellow and the difference between Hive and Spark as seen in the following section.

\subsubsection{CPU utilization for Query 5 at 1TB example}

\begin{figure*}[]
	\centering
	\subfigure[Hive-on-Tez + MLlib2]{
		\includegraphics[width=0.9\textwidth]{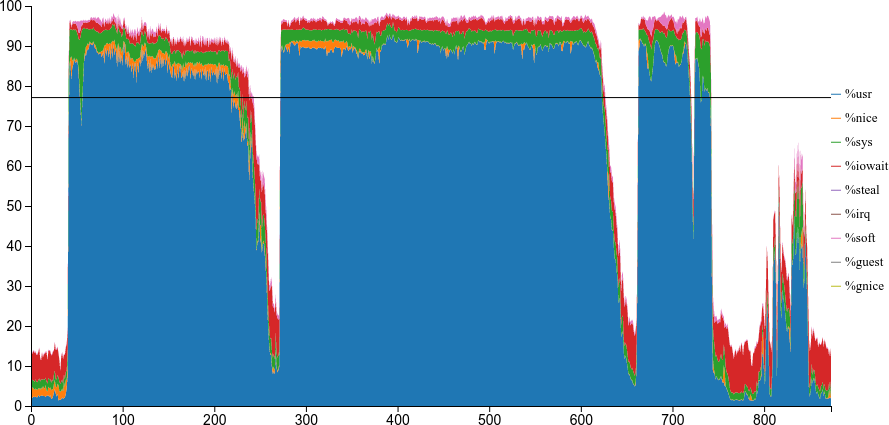}
		\label{fig:Q5_Tez}		
	}
	\\
	\subfigure[Spark 2.1 + MLlib2]{
		\includegraphics[width=0.9\textwidth]{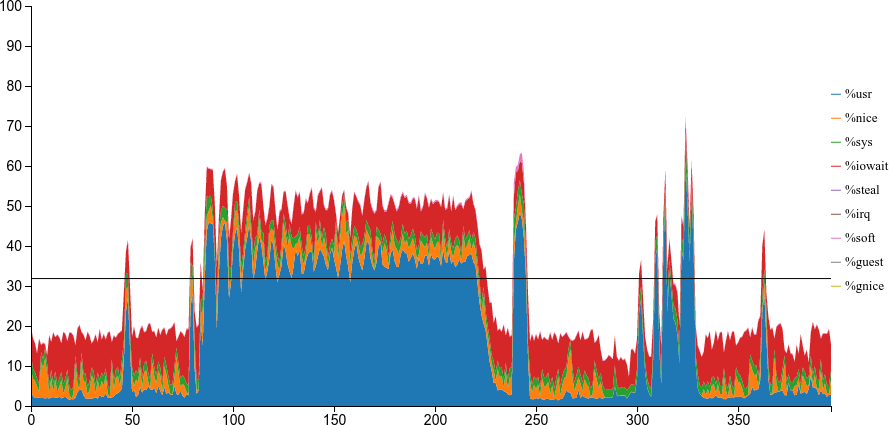}
		\label{fig:Q5_Spark}		
	}
	\caption{CPU\% utilization over time (s.) comparison for Hive and Spark for query 5 at 1TB}
	\label{fig:Q5}
\end{figure*}

Figures~\ref{fig:Q5_Tez} and ~\ref{fig:Q5_Spark} shows the average CPU utilization percentage for Hive-on-Tez and Spark respectively for query 5 on HDI.  Query 5 was chosen as it was found to have the highest resource requirements of the Machine Learning queries.
For this query, Spark takes less than half the time as Hive.  This is interesting as both are using the exact MLlib v2 library.  While Tez is CPU bound in general, we can see that for Spark, the main bottleneck is the I/O wait.  Taking a closer look into disk and network utilization, it was found that the I/O wait in Spark is caused by the network, as the disk requirements are much lower than Hive's.  For this query, Spark is more optimal on the storage subsystem.  While this is an example, we are currently working in more in-depth comparison of the engines and queries.  Especially, as lower query times can lead to differentiated resource bottlenecks, \ie CPU I/O wait.

\section{Power tests from 1GB to 10TB}

This section presents the execution times of the 3 providers when increasing the data scale factor from 1GB by factors of ten up to 10TB.   
The main comparison is centered at the 1TB scale, while a brief comparison is made for all the scale factors for all of the queries. 
Finally, the results at 10TB for the SQL-queries is presented, and the errors found on the scaling process.

\subsubsection{Execution times at 1TB}
\label{sec:times}

\begin{figure}[]
	\centering
	\includegraphics[width=1.0\textwidth]{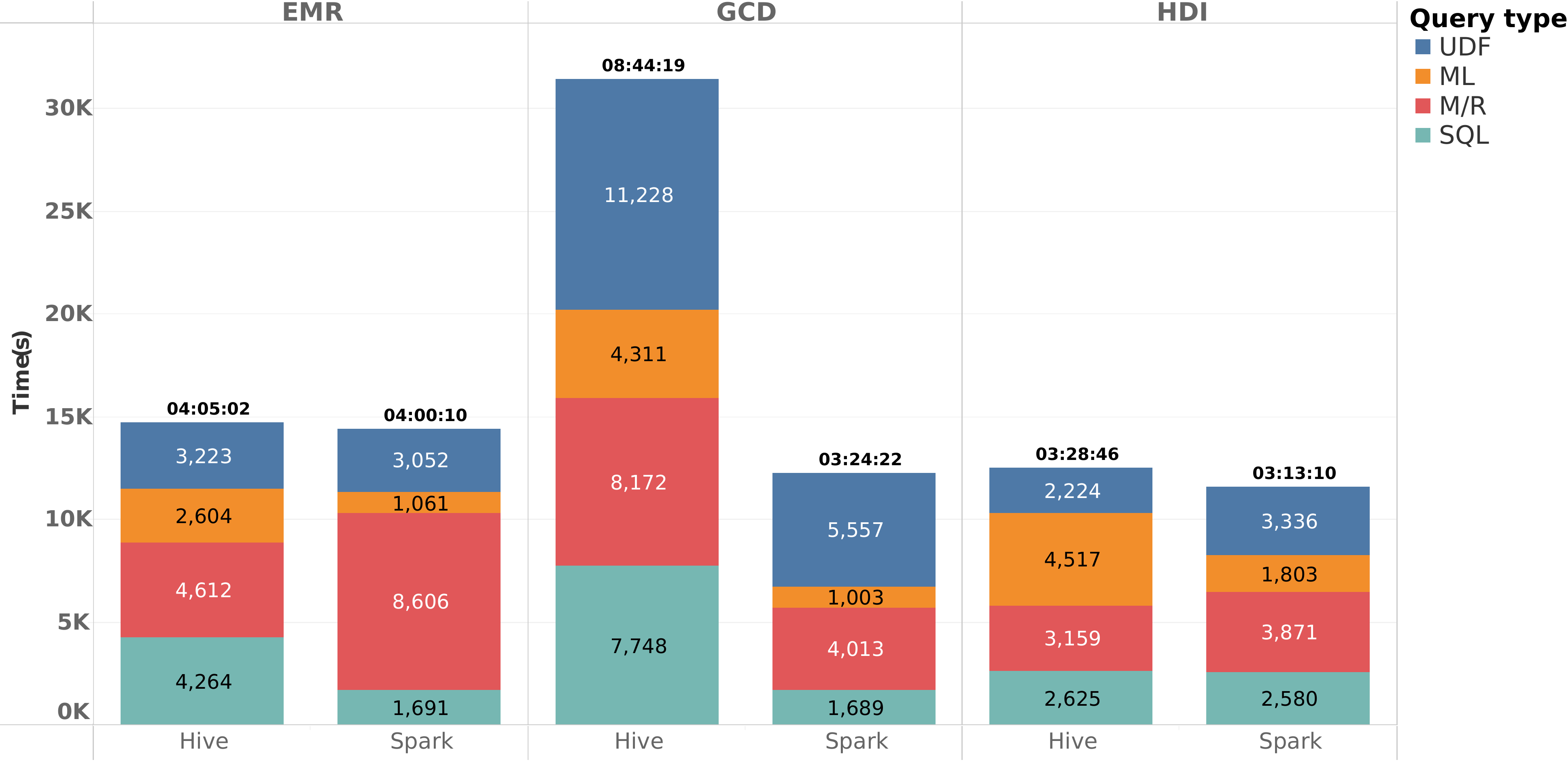}
	\caption{Execution times for a power test at 1TB by provider and framework (Hive vs. Spark)}
	\label{fig:times_1TB}
\end{figure}

Figure~\ref{fig:times_1TB} presents the total time for the \emph{power runs} at 1TB for both Hive and Spark v2 for the different providers in a bar chart.  
Each bar is subdivided into the different query categories: SQL, M/R, ML, and UDF.  
The Y-axis shows the number of seconds employed by each section and in total.
Results are separated by provider first,  EMR, GCD, and HDI;  internally by Hive and Spark respectively.

Results show that Spark is slightly faster than Hive in both HDI ---which gets the fastest results--- and EMR.  
In GCD, Spark gets the second best result while the Hive results is more than two times the slowest result across providers taking more than 8 hours for the full run.
The reason for Hive in GCD being at least twice as slow is due to GCD not using Tez as the execution engine, and using the legacy Map/Reduce engine.  This result is consistent with previous work~\cite{BD16}.

Besides the GCD Hive result, the rest take between 3 and 4 hours for the full run.  While results look similar, especially when comparing Hive vs. Spark in EMR and HDI, times for the different categories of queries differ significantly.
In EMR, only the UDF queries take similar times, while Spark is more than twice as fast in the SQL queries, but almost twice as slow in the M/R ones.  For the ML queries, in all providers Spark gets the best times, being at least twice as fast than in Hive.
Another result that highlights it the long execution time for the M/R portion in EMR with Spark, taking more than twice the time in the rest of the providers.  Having a closer look at the M/R portion, it is query 2 the one that takes most of the time in EMR.   

The difference across providers by query type, shows that tuning the configuration from the defaults can lead to significant improvements in execution times.
In this case, while HDI gets the best times for both Hive and Spark, the ML queries in EMR with Hive take about half the time, while  GCD's ML is the fastest with Spark.
In the case of query 2 for EMR, in a later experiment increasing Sparks memory, the resulting times where then similar to both HDI and GCD.
   
\subsubsection{Errors found during the 1TB tests}
\label{sec:errors}

Up to 1TB, everything was run with the out-of-the-box configuration in Hive, except for some queries in Spark at 1TB:
\begin{itemize}
	\item Queries 14 and 17 (SQL) requires cross joins to be enabled in Spark v2.  A feature which is disabled by default, and it was using the default value in the 3 providers.
	\item Queries 2 and 30 required more memory overhead than the default (384MB) and containers were being killed.  The increased setting was:\\ \emph{spark.yarn.executor.memoryOverhead}.
	\item Queries 3, 4, and 8 required more executor memory, and where failing with \emph{TimSort java.lang.OutOfMemoryError: Java heap space at org.apache.spark.util. collection.unsafe.sort.UnsafeSortDataFormat.allocate}. The increased setting was: \emph{spark.yarn.executor.memory}.
\end{itemize}

\subsubsection{Scalability up to 1TB}

\begin{figure}[]
	\centering
	\includegraphics[width=1.1\textwidth]{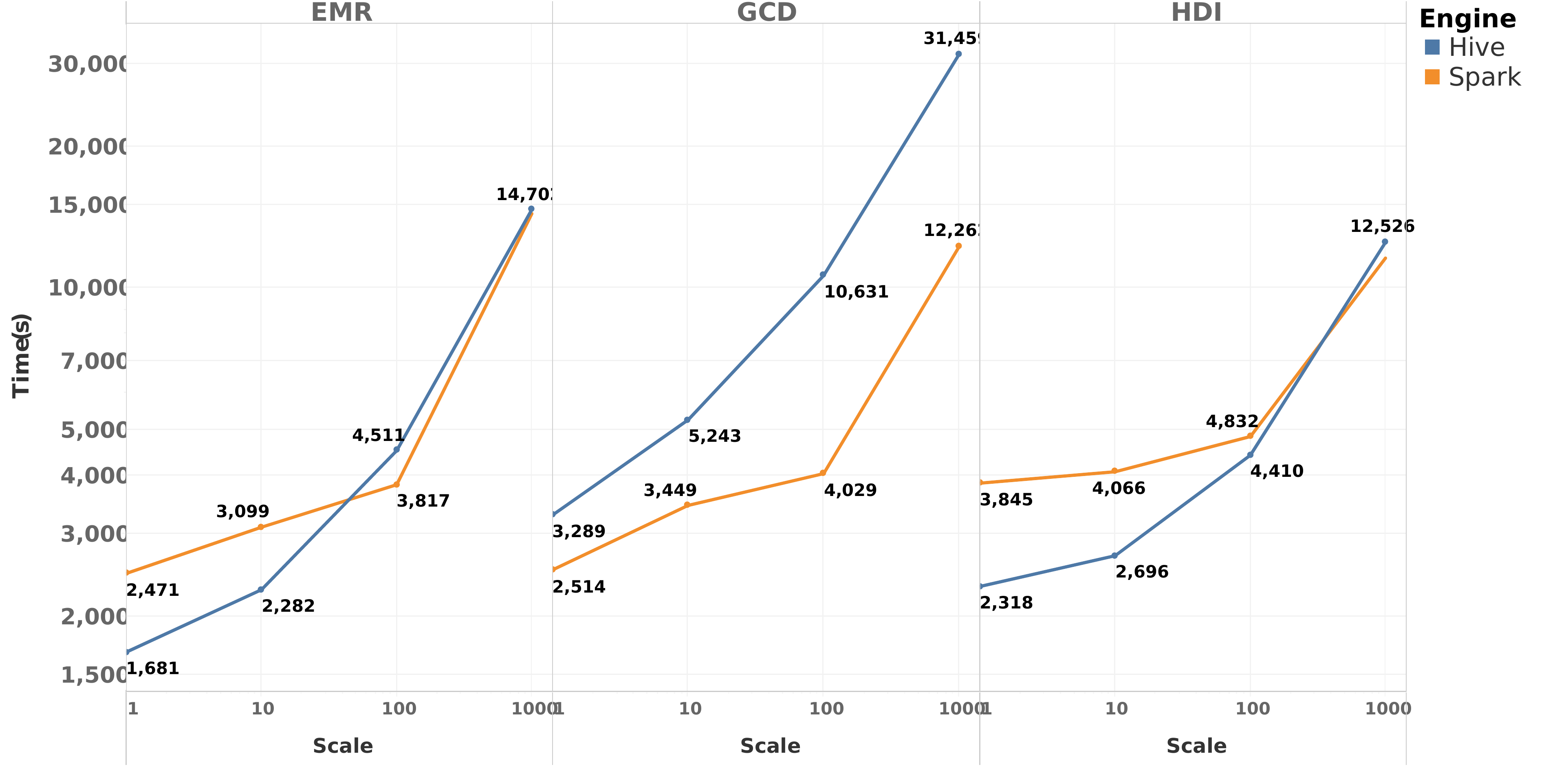}
	\caption{Scalability from 1GB to 1TB for Hive and Spark by provider. Log-log scale.}
	\label{fig:scale_1TB}
\end{figure}

Figure~\ref{fig:scale_1TB} presents the scalability and times for both Hive and Spark for each provider from 1GB to 1TB.  Note that both axes are in log scale for visualization purposes of the different scales.
One thing to note ---excluding GCD--- is that Hive is significantly faster than Spark at the smaller scales.  While from 100GB, Spark is very close to Hive in both EMR and HDI.
This similarity on results could also mean that a common bottleneck is reached by both frameworks, and it is part of our ongoing study.

\subsection{Execution times at 10TB (SQL-only)}
\label{sec:times10}

\begin{figure}[]
	\centering
	\includegraphics[width=1.0\textwidth]{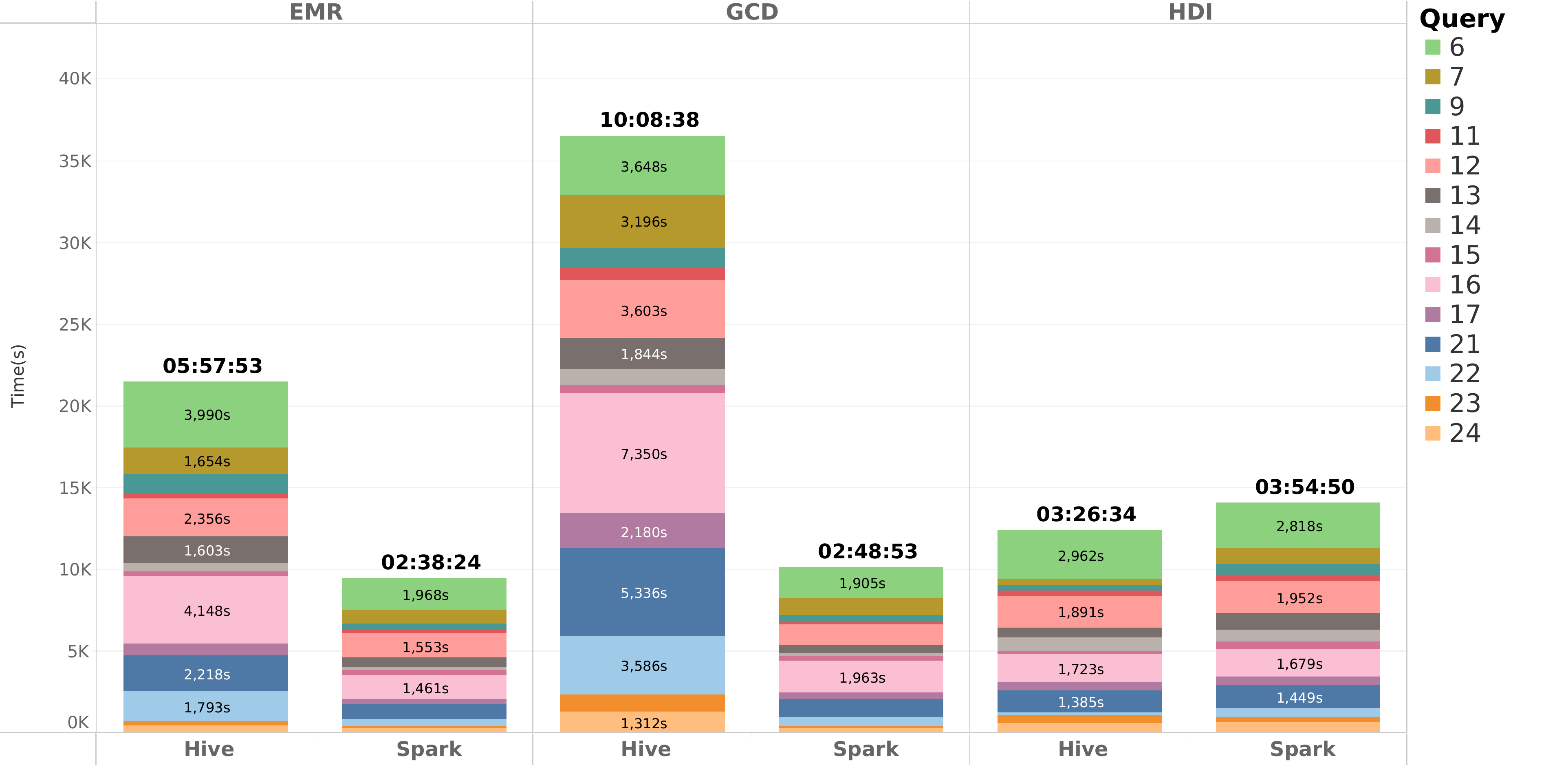}
	\caption{Execution times of SQL-only queries (14) at 10TB by provider and framework}
	\label{fig:times_10TB}
\end{figure}

Figure~\ref{fig:times_10TB} presents the results by query for only the declarative set of the queries ---SQL-only--- for both Hive and Spark.  The reason for presenting the SQL-only queries is due to errors in query execution.  While with some trial-and-error tuning we were able to run all of the queries in Hive in HDI and EMR, we were not able with this cluster sizes to run it successfully with Spark v2 across providers and Hive (M/R) in GCD.  10TB seem to be the limits for the current cluster setup.  However, with the SQL-only queries, we can already see some differences from 1TB and across providers.

Total duration for the queries is between two and a half to ten hours.  Results at 10TB are in proportion to the 1TB results if we only look at this type of queries in Figure~\ref{fig:times_1TB}.  In  the 10TB case in Figure~\ref{fig:times_10TB}, again GCD with Hive is by far the slowest of the systems, while with Spark obtains the second best time.  
EMR with Spark is the fastest also at 10TB for the SQL only, being twice as fast as with Hive.  HDI results are similiar in proportion to the 1TB results compared to the rest.

At 10TB, the memory increase for Spark at 1TB was also needed.  As well as an extra time out setting in HDI for queries 14 and 17 (cross-joins).  Cross joins in Spark 2 are disabled as they are not efficient at the moment.  The updated setting was: \emph spark.sql.broadcastTimeout (default 300). 
We are still analyzing the rest of the errors at 10TB, but it is out of the scope of this study.

\subsubsection{Scalability up to 10TB (SQL-only)}

\begin{figure}[]
	\centering
	\includegraphics[width=1.0\textwidth]{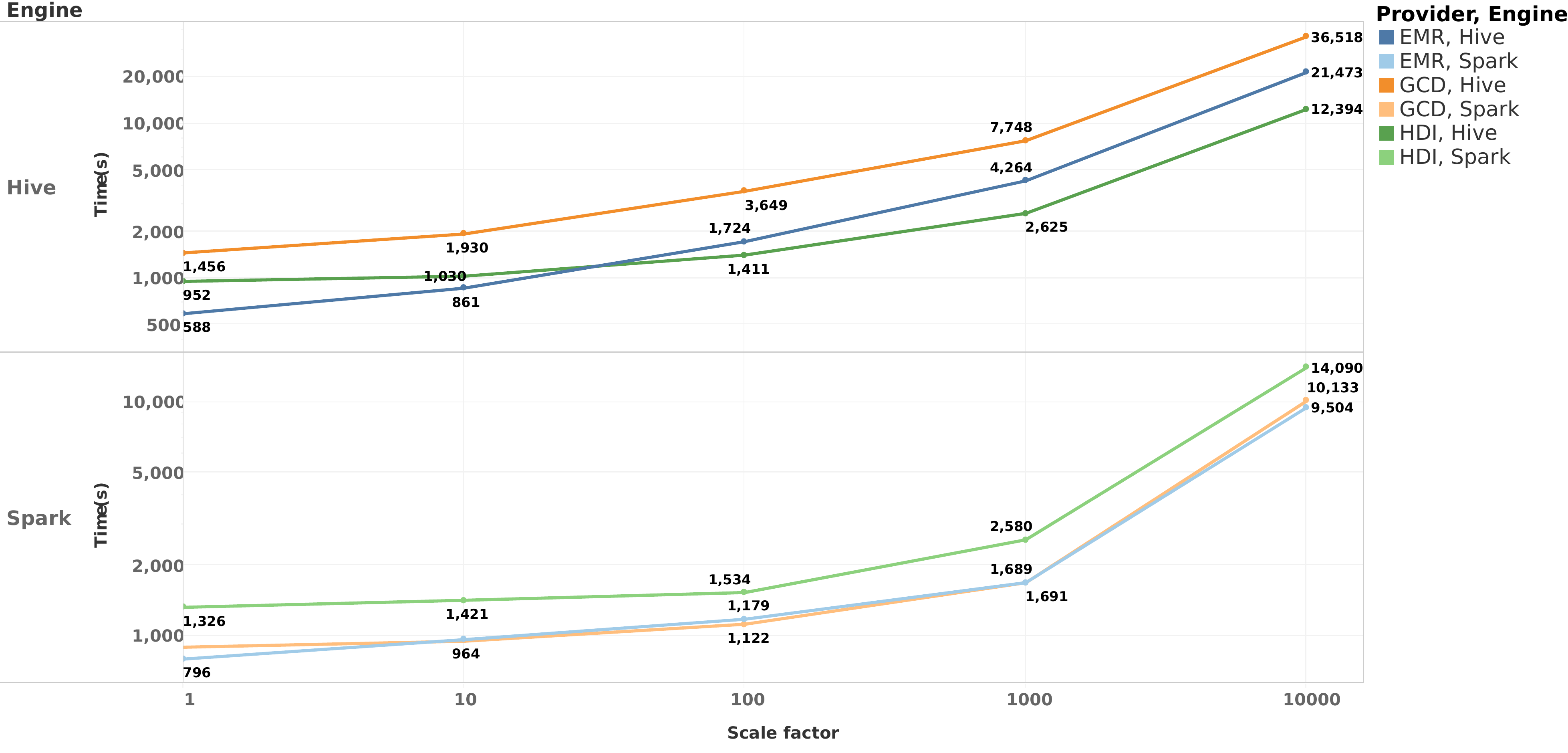}
	\caption{Scalability from 1GB to 10TB for Hive and Spark by provider (SQL-only queries). Log-log scale.}
	\label{fig:scale_10TB}
\end{figure}

Figure~\ref{fig:scale_10TB} presents the scalability and times for both Hive and Spark for each provider from 1TB to 10TB only for the 14 SQL queries.  Note that both axes are in log scale for visualization purposes of the different scales.
It is interesting to see in the EMR case that Spark is twice as fast than Hive at 1 and 10TB.  While on HDI, they obtain similar results.

\section{Additional experiments}
\label{sec:additional}

This sections briefly adds information on additional testing that was done with different versions of the frameworks.  

\subsubsection{Spark 2.0.2 vs 2.1.0 on GCD 1GB-1TB}

On GCD, we have also tested their sofware versions 1.1 to the preview version.  Here we could see the difference of Spark 2.0.2 to Spark 2.1.0 on exactly the same hardware.  We found that Spark 2.1 is a bit faster at small scales, but slower at 100GB and 1TB specifically on the UDF/NLP queries.

\subsubsection{Spark 1.6.3 vs 2.1.0 MLlib 1 vs 2.1 MLlib 2 on HDI 1GB-1TB}

In HDI which uses the HDP distribution we could test Spark version 1.6.3 and Spark 2.1 on exactly the same cluster.  We found that Spark 2.1 is always faster than 1.6.3 this case.
In the HDI cluster we also compared MLlib v1 to v2.  MLib v2 makes use of the newer dataframes API, opposed to RDDs in v1.  We found v2 to be only slightly faster than V1.

\subsubsection{Throughput runs} 

\begin{figure}[]
	\centering
	\includegraphics[width=1.0\textwidth]{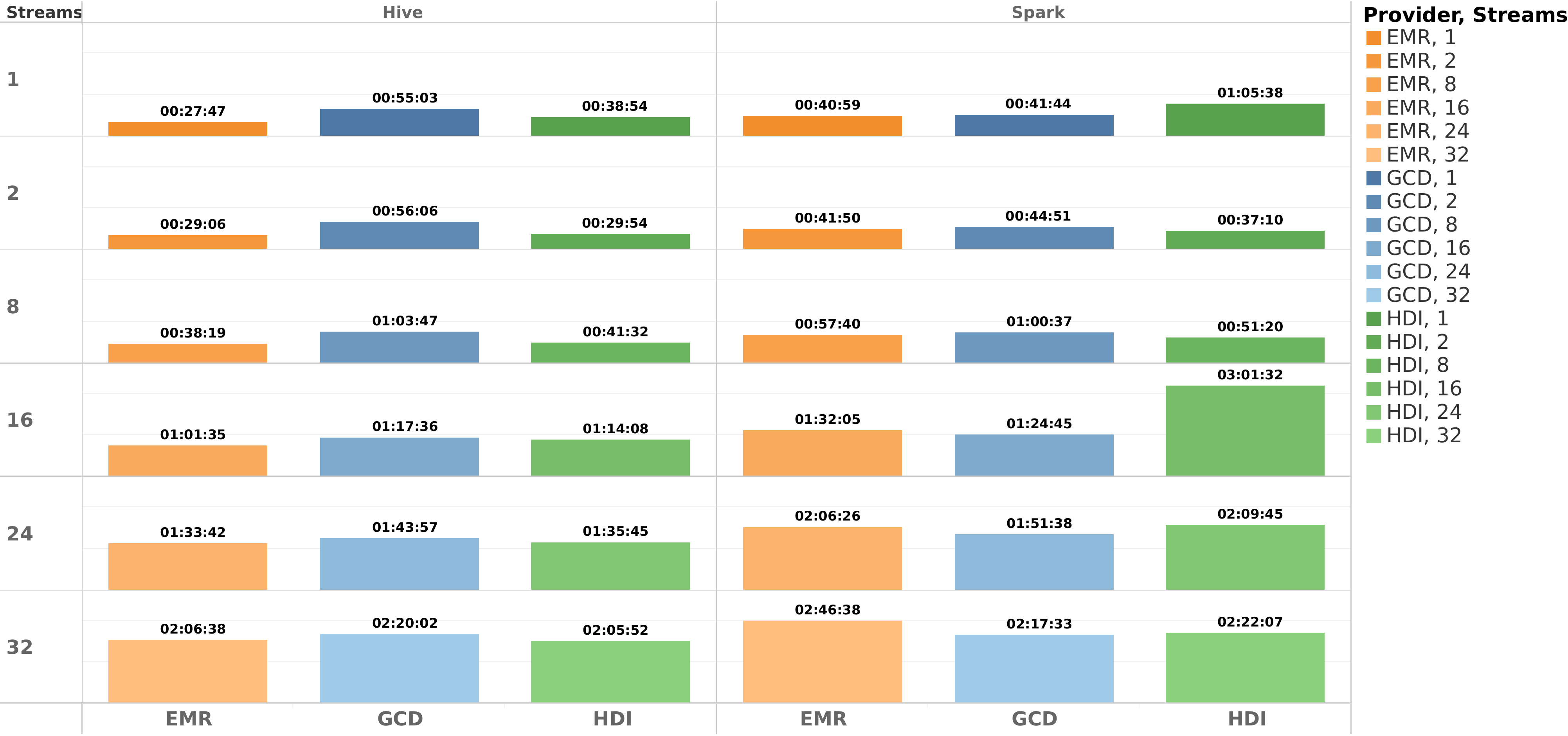}
	\caption{Throughput runs from 1 to 32 streams at 1GB scale in Hive and Spark by provider}
	\label{fig:through}
\end{figure}

Figure~\ref{fig:through} shows the results for Hive and Spark for each provider as the number of concurrent streams (clients) are increased. Streams are increased from 1 (no concurrency) to 32 streams.
At 32 streams we can se that the best numbers are obtained by Hive in HDI and EMR.  We can also see a great variation of results in HDI with Spark, as with 16 streams is the slowest of the systems, but the fastest at 32 streams.  This situation also highlights the variability of cloud results as we have studied previously in~\cite{BD16}.

\section{Conclusions}

This study presented first a characterization of the resource consumption by each BigBench query. As BigBench is a recent benchmark with constant changes to the implementations, there is still little knowledge about the difference in queries.  Such knowledge can be useful to quickly benchmark a new SUT by \emph{cherry-picking} queries that stresses different resources. Also, it is useful to validate results and compare future implementations.  
In particular, we have found that the M/R type queries utilizes the most resources, with query 2 having the highest utilization of all.  For the UDF queries, query 30 is by far the most resource hungry, having the highest network and disk requirements.  On the ML queries, query 5 has the highest resource requirements, while query 6 from the SQL-only group.

While Hive-on-Tez in general uses a \emph{thin} container strategy as in classical M/R and Spark uses \emph{fat} containers, results show that more than half of the queries share similar resource requirements. The SQL-only queries were the ones with more differences between the frameworks.  On our preliminary tests we have also found that Hive-on-Tez improves the performance up to 4x over Hive-on-MapReduce as used in GCD.  Spark MLlib has also improved the performance over the Mahout Machine Learning library, and MLlib v2 over v1 using the dataframes API in a moderate amount.  The first BigBench implementation was for both M/R and Mahout can now be considered legacy and should be avoided. 
Spark 2.1 is faster than previous versions, however, Spark improvements are within the 30\% range, and was not found to be as pronounced as in Web articles.
Hive-on-Tez (+ MLlib for ML) are still faster than Spark at lower scales, but this difference narrows down at larger scales.  We are currently investigating if due to a common hardware bottleneck of framework at scale, but Spark shows improved performance under concurrency

Performance was found to be similar among providers for the tested configurations.
All providers currently have up to date (2.1.0) and well tuned versions of Spark.  This is contrast of our previous study using a TPC-H benchmark the previous year~\cite{BD16}.
All providers using medium-sized, 128-core clusters could run BigBench up to 1TB out-of-the-box with minimal memory tuning on Spark.  While at 10TB, queries start failing and only could complete the SQL-only queries for both Hive and Spark.
While BigBench is a recent benchmark, it can already help us guide our decision making in Cloud providers, Big Data frameworks, and Machine Learning libraries.
However, it still needs more engines to be added to the public implementation and more results available.


\section*{Acknowledgements}  
This project has received funding from the European Research Council (ERC) under the European Union’s Horizon 2020 research and innovation programme (grant agreement No 639595). It is also partially supported by the Ministry of Economy of Spain under contract TIN2015-65316-P and Generalitat de Catalunya under contract 2014SGR1051, by the ICREA Academia program, and by the BSC-CNS Severo Ochoa program (SEV-2015-0493).

\bibliographystyle{abbrv}
\bibliography{aloja}

\begin{thebibliography}{10}

\bibitem{Boncz2014}
P.~Boncz, T.~Neumann, and O.~Erling.
\newblock Tpc-h analyzed: Hidden messages and lessons learned from an
  influential benchmark.
\newblock {\em Performance Characterization and Benchmarking, TPCTC 2013}.

\bibitem{BigBench2TPCxBB}
P.~Cao, B.~Gowda, S.~Lakshmi, C.~Narasimhadevara, P.~Nguyen, J.~Poelman,
  M.~Poess, and T.~Rabl.
\newblock {\em From BigBench to TPCx-BB: Standardization of a Big Data
  Benchmark}.

\bibitem{Floratou2015}
A.~Floratou et~al.
\newblock Benchmarking sql-on-hadoop systems: Tpc or not tpc?
\newblock {\em Big Data Benchmarking: 5th International Workshop, WBDB 2014,
  Germany, 2014}.

\bibitem{Floratou2014}
A.~Floratou, U.~F. Minhas, and F.~\"{O}zcan.
\newblock Sql-on-hadoop: Full circle back to shared-nothing database
  architectures.
\newblock {\em Proc. VLDB Endow.}, 2014.

\bibitem{BigBench}
A.~Ghazal, T.~Rabl, M.~Hu, F.~Raab, M.~Poess, A.~Crolotte, and H.-A. Jacobsen.
\newblock Bigbench: Towards an industry standard benchmark for big data
  analytics.
\newblock In {\em Proceedings of the 2013 ACM SIGMOD International Conference
  on Management of Data}, SIGMOD '13, pages 1197--1208, New York, NY, USA,
  2013. ACM.

\bibitem{SPECBD}
S.~R. B. D.~W. Group.
\newblock
  \url{https://research.spec.org/working-groups/big-data-working-group.html},
  2016.

\bibitem{HDP}
H.~D.~P. (HDP).
\newblock \url{http://hortonworks.com/products/hdp/}, 2016.

\bibitem{Hive}
A.~Hive.
\newblock https://hive.apache.org/, 2016.

\bibitem{HiBench}
S.~Huang et~al.
\newblock {The HiBench benchmark suite: Characterization of the MapReduce-based
  data analysis}.
\newblock {\em Data Engineering Workshops, 22nd Int. Conf. on}, 2010.

\bibitem{BigBench_github}
Intel.
\newblock Big-data-benchmark-for-big-bench:
  \url{https://github.com/intel-hadoop/Big-Data-Benchmark-for-Big-Bench}, 2016.

\bibitem{D2F}
T.~Ivanov.
\newblock D2f tpc-h benchmark repository:
  https://github.com/t-ivanov/d2f-bench, 2016.

\bibitem{Ivanov2016}
T.~Ivanov and M.-G. Beer.
\newblock {\em Performance Evaluation of Spark SQL Using BigBench}.

\bibitem{ForresterPaaS}
M.~G. Noel~Yuhanna.
\newblock Elasticity, automation, and pay-as-you-go compel enterprise adoption
  of hadoop in the cloud.
\newblock {\em The Forrester Wave: Big Data Hadoop Cloud Solutions}, Q2, 2016.

\bibitem{Pavlo2009}
A.~Pavlo, E.~Paulson, A.~Rasin, D.~J. Abadi, D.~J. DeWitt, S.~Madden, and
  M.~Stonebraker.
\newblock {A Comparison of Approaches to Large-Scale Data Analysis}.
\newblock In {\em SIGMOD}, pages 165--178, 2009.

\bibitem{BD15}
N.~Poggi, J.~L. Berral, D.~Carrera, N.~Vujic, D.~Green, J.~Blakeley, et~al.
\newblock From performance profiling to predictive analytics while evaluating
  hadoop cost-efficiency in aloja.
\newblock In {\em Big Data (Big Data), 2015 IEEE International Conference on},
  2015.

\bibitem{BD16}
N.~Poggi, J.~L. Berral, T.~Fenech, D.~Carrera, J.~Blakeley, U.~F. Minhas, and
  N.~Vujic.
\newblock The state of sql-on-hadoop in the cloud.
\newblock In {\em 2016 IEEE International Conference on Big Data (Big Data)},
  pages 1432--1443, Dec 2016.

\bibitem{BD14}
N.~Poggi, D.~Carrera, N.~Vujic, J.~Blakeley, et~al.
\newblock {ALOJA:} {A} systematic study of hadoop deployment variables to
  enable automated characterization of cost-effectiveness.
\newblock In {\em 2014 {IEEE} Intl. Conf. on Big Data, Big Data 2014,
  Washington, DC, USA, October 27-30, 2014}.

\bibitem{HSEU2017EU}
N.~Poggi and A.~Montero.
\newblock Using bigbench to compare hive and spark versions and features.

\bibitem{Compendium}
M.~P. A. Q. J. P. N. P. J.~B. Todor~Ivanov, Tilmann~Rabl.
\newblock Big data benchmark compendium.
\newblock {\em 7th TPC Technology Conference on Performance Evaluation and
  Benchmarking (TPCTC 2015) USA}.

\bibitem{BigBench_submissions}
TPC.
\newblock Tpcx-bb official submissions:
  \url{http://www.tpc.org/tpcx-bb/results/tpcxbb_perf_results.asp}, 2016.

\bibitem{TPC-H}
{Transaction Processing Performance Council}.
\newblock {TPC Benchmark H - Standard Specification}, 2014.
\newblock Version 2.17.1.

\bibitem{TPC-DS}
{Transaction Processing Performance Council}.
\newblock {TPC Benchmark DS - Standard Specification}, 2015.
\newblock Version 1.3.1.

\bibitem{SruthiEMR}
S.~Vijayakumar.
\newblock Hadoop based data intensive computation on iaas cloud platforms.
\newblock {\em UNF Theses and Dissertations}, page Paper 567, 2015.

\bibitem{Yahoo}
T.~Yahoo Betting~on Apache~Hive and YARN.
\newblock
  \url{https://yahoodevelopers.tumblr.com/post/85930551108/yahoo-betting-on-apache-hive-tez-and-yarn},
  2014.

\bibitem{Spark}
M.~Zaharia, R.~S. Xin, P.~Wendell, T.~Das, M.~Armbrust, A.~Dave, X.~Meng,
  J.~Rosen, S.~Venkataraman, M.~J. Franklin, A.~Ghodsi, J.~Gonzalez,
  S.~Shenker, and I.~Stoica.
\newblock Apache spark: A unified engine for big data processing.
\newblock {\em Commun. ACM}.

\bibitem{Zhang2014cloud}
Z.~Zhang, L.~Cherkasova, and B.~T. Loo.
\newblock Exploiting cloud heterogeneity for optimized cost/performance
  mapreduce processing.
\newblock In {\em CloudDP 2014}.

\bibitem{zhang2014optimizing}
Z.~Zhang et~al.
\newblock Optimizing cost and performance trade-offs for mapreduce job
  processing in the cloud.
\newblock In {\em NOMS 2014}.

\end{thebibliography}

\end{document}